\documentstyle[12pt]{article}
\title{On evaluation of two-loop self-energy diagram \\
with three propagators}

\author{F. A. Lunev \thanks{e-mail address:
lunev@hep.phys.msu.su } \\ {\em Physical Department, Moscow
State University, Moscow, 119899, Russia} }

\date{ \ \ \  }

\begin{document}
\baselineskip 24pt

\maketitle

\begin{abstract}
Small momentum expansion of the "sunset" diagram with
three different masses is obtained. Coefficients at powers of
$p^2$ are evaluated explicitly in terms of dilogarithms and
elementary functions.  Also  some power expansions of
"sunset" diagram in terms of different sets of variables are
given.  \end{abstract}

\vspace{1.5cm}

The "sunset" diagram (see Fig. 1) was the object of investigation
in several recent works \cite{DT,DTS,BBBS,BT,GB}. Such activity
is initiated, certainly, by rather high precision of experimental
testing of the standard model that needs sometimes evaluation of
two-loop Feynman diagrams for comparison of the theory and
experiments.

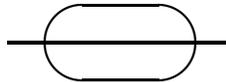
\begin{figure}[h]
\unitlength=5mm
\centering
\thicklines
\begin{picture}(8,5)(0,2)
\put(1,4){\line(1,0){6}}
\put(4,4){\oval(4,2)}
\put(6,4){\line(1,0){1}}
\end{picture}
\caption{The "sunset" diagram}
\end{figure}

\vskip 0.5cm

Probably, the "sunset" diagram cannot be expressed through known
special functions of one variable. So for investigation of this
diagram it was used numerical methods \cite{BT,GB} and some
approximations schemes, such as expansions in powers of $p^2$
(and possibly logarithms) \cite{DT,DTS,BBBS} \footnote{See also
an old paper by Mendels \cite{M} where it was obtained expansion
of "sunset" diagram in equal mass case in powers of $p^2/p^2 +
m^2$}.

In particular, in paper \cite{BBBS} it was obtained small
momentum expansion of "sunset" diagram in terms of Lauricella
series. These series are convergent in the region

\begin{equation}
\sqrt{p^2} + m_1 +m_2 < m_3 \label{(1)}
\end{equation}

\noindent where $m_i$, $i=1,2,3$, are masses in propogators. But
for majority of applications $\hbox{$m_1 + m_2 > m_3$}$ for any
numbering of masses. Such situation takes place, for instance,
in the course of  evaluation of self-energy of Higgs boson (when
$m_1 = m_2 = m_3 = m_H$), and in the course of evaluation of
contribution to $Z$-boson self-energy due to vertex $Z^2 W W^* $
(when $m_1 = m_2 = m_W, \ \ m_3 = m_Z $).

The aim of the present paper is to complete the results of
\cite{BBBS} and to obtain small momentum expansion of "sunset "
diagram that can be used for all values of masses, and to
evaluate in closed form coefficients at powers of $p^2$. In
addition, we will give some power expansions of "sunset" diagram
that are convergent, at least, in the region

\begin{equation}
|p^2| + m^2_3 < (m_1 + m_2 )^2 \label{(2)}
\end{equation}

\noindent Obviously, these expansions can be used for all values
of masses if one defines $m_3$ as $ \hbox{min} \{ m_1, m_2, m_3
\} $.

Let $I = I (p^2, m^2_1, m^2_2, m^2_3 ) $ be the "sunset"
diagram.  Our main tool will be the following one-fold integral
representation for the function $I$:

$$
I = \int ^{\infty}_{(m_1 + m_2)^2 } d \sigma ^2
\rho(\sigma ^2, m_1, m_2 )
\left\{ J (p^2, m_3^2, \sigma ^2) \right.
$$

\begin{equation}
\left.
 - \pi ^2 \left[
\frac{m^2_3}{\sigma ^2} \log \frac{m^2_3}{\sigma ^2} + m^2_3 f_1
(\sigma ^2) +p^2 f_2(\sigma ^2) + f_3 (\sigma ^2)  \right]
\right\} \label{(3)}
\end{equation}

\noindent where $J$ is one-loop "bubble" diagram (see Fig.2),
$\rho$ is spectral function of $J$,

\begin{equation}
\rho = \pi ^2 \sqrt
 { \left( 1 - \frac{(m_1 + m_2)^2}{ \sigma ^2} \right)
 \left( 1 - \frac{(m_1 - m_2)^2}{ \sigma ^2} \right) }
 \label{(4)}
\end{equation}

\noindent and functions $f_i, \ \ i=1,2,3$ must be defined in
such a way that integral (\ref{(3)}) converges.

\begin{figure}[h]
\unitlength=5mm
\centering
\thicklines
\begin{picture}(8,5)(0,2)
\put(1,4){\line(1,0){1}}
\put(4,4){\oval(4,2)}
\put(6,4){\line(1,0){1}}
\end{picture}
\caption{The "bubble" diagram}
\end{figure}
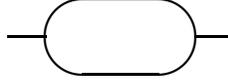

Representation (\ref{(3)}) was obtained in a previous author's
work \cite{L} \footnote{In \cite{L} representation (\ref{(3)})
was derived only for the case $m_1 = m_2$, but, in fact, this
condition was never used and for general case representation
(\ref{(3)}) can be obtained in the same way.}. Very similar
integral representation was obtained independently in above
mentioned papers \cite{BBBS,BT}. Another one-fold integral
representation was given in \cite{GB} .An  analogous one-fold
integral representations for the five propogator self-energy
diagram was derived in \cite{Br,BrFT}.

One notes that, due to renormalization freedom, function $I$ is
defined up to polynomial of the first degree with respect to $p^2$.
So it is sufficient to evaluate the function

\begin{equation}
\bar {I} (p^2,...) = I (p^2,...)  - I (0,...) - p^2 \frac{dI}{dp^2}
(0,...) \label{(5)}
\end{equation}

Comparing (\ref{(3)}) and (\ref{(5)}), one obtains the following
integral representation for $\bar I$:

\begin{equation}
\bar I = \int ^{\infty}_{(m_1 + m_2)^2 } d \sigma ^2
\rho(\sigma ^2, m_1, m_2 )
\bar J (p^2, \sigma ^2, m_3^2) \label{(6)}
\end{equation}

\noindent where

\begin{equation}
\bar {J} (p^2,...) = J (p^2,...)  - J (0,...) - p^2 \frac{dJ}{dp^2}
(0,...) \label{(7)}
\end{equation}

The function $\bar J $ can be represented by the following power
series:

\begin{equation}
\bar J = \pi ^2
\sum \limits _{k \geq 2, \, l \geq 0}
\frac{(k + l - 1)! (k + l)!}{l! (2k + l +1)!}
\left( - \frac {p^2}{\sigma ^2} \right) ^k
\left( 1 - \frac {m^2_3}{\sigma ^2} \right) ^l  \label{(8)}
\end{equation}

The proof of (\ref{(8)}) is very simple \footnote{Another proof
was given in \cite{D}.}. One considers the representation of $J$
through Feynman parameters:

\begin{equation}
J = - \pi ^2 \int _0^1 d \xi \log \left[
 \frac{p^2 \xi (1 - \xi) + \xi m_3^2
+ (1 - \xi) \sigma ^2}{\sigma ^2}
 \right]  \label{(9)}
\end{equation}

\noindent Expanding the integrand in powers series with respect
to

 $$
 \frac{p^2 \xi (1 - \xi) +
\xi (m_3^2 - \sigma ^2)}{\sigma ^2}
 $$

\noindent and integrating, one obtains (\ref{(8)}). These
operations are correct if

\begin{equation}
\max _{0 \leq \xi \leq 1} \left|
 \frac{p^2 \xi (1 - \xi) +
\xi (m_3^2 - \sigma ^2)}{\sigma ^2}
\right|  \leq 1         \label{(10)}
\end{equation}

\noindent From (\ref{(10)}) one can derive that series (\ref{(8)})
converges, at least, if

\begin{equation}
|p^2 | + m^2_3 < \sigma ^2
\label{(11)}
\end{equation}

One introduces dimensionless variables

$$
x= \frac{(m_1 + m_2)^2}{\sigma ^2}, \ \ \ \ y= \frac{m_3^2}{(m_1
+ m_2)^2} < 1,
$$

\begin{equation}
z = - \frac{p^2}{(m_1 + m_2)^2}, \ \ \ \ \lambda = \frac{(m_1 -
m_2)^2}{(m_1 + m_2)^2} \label{(12)}
\end{equation}

In these variables, using definition of Gauss hypergeometric
function ${ \ }_2 F_1$, one can rewrite (\ref{(8)}) in the
following way:

\begin{equation}
\bar J = \pi ^2 \sum \limits _{k \geq 2} x^k z^k
\frac{(k-1)!k!}{(2k+1)!} { \ }_2 F_1 (k, \ k+1; \ 2k+2; \ 1-xy)
\label{(13)}
\end{equation}

\noindent But

\begin{equation}
\frac{(k-1)!k!}{(2k+1)!} { \ }_2 F_1 (k, \ k+1; \ 2k+2; \ u) =
\frac{1}{k! (k+1)! } \frac{d^k}{du^k}
(1-u)^{k+1}\frac{d^k}{du^k} \frac{\log (1 -u)}{u} \label{(14)}
\end{equation}

\noindent (see, for instance, \cite{E}). From the last formula,
after some elementary transformations, one can obtain:

$$
\frac{(k-1)!k!}{(2k+1)!} { \ }_2 F_1 (k, \ k+1; \ 2k+2; \ 1-xy)
$$
\begin{equation}
= \frac{1}{k!(k+1)!} \frac{1}{x^{k-2}} \frac{d}{dx}
\frac{d^k}{dy^k} y^{k+1} \frac{d^{k-1}}{dy^{k-1}} \frac{\log
(xy)}{y(1-xy)} \label{(15)}
\end{equation}

Substituting (\ref{(13)}) and (\ref{(15)}) in (\ref{(6)}) and
integrating by parts with respect to $x$, one finds:

\begin{equation}
\bar I = \pi ^4 (m_1 + m_2 )^2 \sum \limits _{k \geq 2}
\frac{z^k}{k!(k+1)!} f_k(y, \lambda)   \label{(16)}
\end{equation}

\noindent where functions $f_k (y, \lambda ) $ are defined by
formulas:

\begin{equation}
f_k(y, \lambda) = - \frac{d^k}{dy^k} y^{k+1}
\frac{d^{k-1}}{dy^{k-1}} \frac{f(y, \lambda )}{y} \label{(17)}
\end{equation}

\begin{equation}
f(y, \lambda ) = \log y + \int _0^1 dx \frac{\log
(xy)}{1-xy} \frac{d}{dx} \sqrt {(1-x)(1- \lambda y)}
\label{(18)} \end{equation}

The integral in (\ref{(18)})  can be easily evaluated and we
obtain for $f(y, \lambda)$ the following explicit expression:

$$
f(y, \lambda) = \log y - \frac{\sqrt {\lambda}}{y} \left[ \log
\left( \frac{1 + \sqrt {\lambda}}{1 - \sqrt {\lambda}} \right)
\log  \left(
\frac{4y}{(1- \sqrt {\lambda})^2}   \right) \right.
$$

$$
\left. - \hbox{Sp} \left( -\frac{2 \sqrt{\lambda}}{1+ \sqrt{
\lambda}} \right)  + \hbox{Sp} \left( - \frac{4 \sqrt
{\lambda}}{(1- \sqrt{\lambda})^2} \right)  - \hbox{Sp} \left( -
\frac{2 \sqrt{\lambda}}{1- \sqrt{\lambda}} \right) \right]
$$

$$
+ \frac{(2 \lambda - y \lambda + y)}{2y \sqrt{(1-y)(\lambda
- y)}} \left[ \log \left( \frac{4y}{(1- \sqrt{\lambda})} \right)
\log \left( \frac{t-c}{1-ct} \right) + \hbox{Sp} \left(
\frac{t - 1}{t- \frac{1}{c} } \right) \right.
$$

\begin{equation}
\left. -\hbox{Sp} \left( \frac{t - 1}{t -c} \right) + \hbox{Sp}
\left( \frac{1-t}{1- \frac{t}{c}} \right) - \hbox{Sp} \left(
\frac{1-t}{1-ct} \right) + \hbox{Sp} \left( \frac{1-t^2}{1-ct}
\right)
- \hbox{Sp} \left( \frac{1-t^2}{1- \frac{t}{c}} \right)
\right] \label{19}
\end{equation}

\noindent where $ \hbox{Sp}(x) $ is the Spence function (or
dilogarithm),

\begin{equation}
t= \frac{1+ \sqrt{ \lambda}}{1- \sqrt{\lambda}}, \ \ \ c=
\frac{1}{y(1- \lambda)} \left( 2 \lambda - y(\lambda +1) + 2
\sqrt{ \lambda (1 -y)(\lambda -y)} \right)  \label{20}
\end{equation}

Formulas (\ref{(16)}), (\ref{(17)}), and (\ref{19}) give
explicit expressions for coefficients at powers of $p^2$ in
Taylor expansion of "sunset" diagram. But formula (\ref{19})
seems too complicated. Fortunately, in almost all applications
$m_1 = m_2$ (that is, $\lambda =0$), and just in this case
formula (\ref{19}) considerably simplifies:

\begin{equation}
f(y,0) = \log \, y + \frac{1}{\sqrt{y(1-y)}} \hbox{Cl}_2 (2
\arcsin \,  \sqrt{y} ) \label{21}
\end{equation}

\noindent where $\hbox{Cl}_2 (\theta)$ is Clausen integral.

In conclusion, we will derive two series expansions for "sunset"
diagram for the case of three different masses, when direct
using of explicit formulas (\ref{(16)}), (\ref{(17)}), and
(\ref{19}) is difficult due to complexity of formula (\ref{19}).

First, one substitutes the decomposition (\ref{(8)}) in
(\ref{(6)}), expands integrand with respect to $\lambda$ and
performs integration. Then, after some elementary
transformations, one obtains the following result:

$$
\bar I = \frac{\pi^4 (p^2)^2}{(m_1 +m_2)^2}
\sum \limits _{j,k,l \geq 0} \left(
\sum \limits _{n \geq 0} \frac{(k+l+n+1)! \, (k+l+n+2)! \,
\Gamma (n+\frac{3}{2})}{n! \, (2k+l+n+5)! \, \Gamma (j+k+l+n+
\frac{5}{2})} \right) $$

\begin{equation}
\times \frac{(j+k+l)! \, \Gamma (j - \frac{1}{2})}{l! \, j!
\Gamma (- \frac{1}{2})} \lambda ^j (1-y)^l z^k  \label{22}
\end{equation}

Above mentioned operations are correct if condition (\ref{(13)})
is valid for all $\sigma ^2 \geq (m_1 +m_2)^2$. This means, that
series  (\ref{22}) converges, at least, at region (\ref{(2)}).

The sum over $n$ can be evaluated in terms of finite sums. But
the corresponding expression is too cumbersome to be useful. So,
if it is desirably to have closed expression for coefficients at
$\lambda ^j y^l z^k$, it is preferable to use another series
expansion for $\bar I$, namely:

$$
\bar I = \pi ^4 (m_1 +m_2)^2 \sum \limits _{j \geq 0, k \geq
2} \frac{z^k \lambda^j}{2j!} \Gamma (j+ \frac{1}{2}) \left\{
\frac{(j+k-2)!}{k(k+1) \, \Gamma (j+k+ \frac{1}{2})}  \right.
$$

\begin{equation}
\left. + \sum \limits _{l \geq 0}
\frac{(k+l)! \, (k+l+1)! \, (j+k+l-1)!}{k! \, (k+1)! \, l!
\, (l+1)! \,  \Gamma (j+k+l+ \frac{3}{2})} \left[ h_{jkl} +\log
\, y \right] y^{l+1} \right\} \label{23} \end{equation}

\noindent where

$$
h_{jkl} = \psi (j+k+l) + \psi (k+l+1) + \psi (k+l+2)
$$

$$
 - \psi (j+k+l+ \frac{3}{2} )- \psi (l+1) - \psi (l+2)
$$

In order to prove (\ref{23}), it is sufficient to write
function ${ }_2 F_1$ in (\ref{(13)}) as power series with
respect to $y$ and  $\log \, y$, to substitute this series in
(\ref{(6)}), and to perform the integration.

These operations again are correct, if condition (11) is valid
for all $\sigma ^2 \geq (m_1 +m_2)^2$. So the region of
convergence of series (\ref{23}) is not less than one defines
by formula (\ref{(2)}).

Large momentum expansion for "sunset" diagram also may be
derived from integral representation (\ref{(3)}). But in this
case results of \cite{BBBS} seem quite exhaustive and so further
investigations of this problem are unnecessarily.

Author is indebted to A.I.Davydychev and B.A.Smirnov for
interesting discussions and comments.

\end{document}